\documentclass[a4paper]{spie}  
\usepackage[]{graphicx}

\title{Gaia in-orbit realignment. Overview and data analysis} 

\author{Alcione Mora\supit{a}\supit{b} and Amir Vosteen\supit{c}
\skiplinehalf
\supit{a}ESA-ESAC Gaia SOC, P.O. Box 78, 28691 Villanueva de la Ca\~{n}ada, Madrid, Spain; \\
\supit{b}Aurora Technology, Crown Business Centre, Heereweg 345, 2161 CA Lisse, The Netherlands; \\
\supit{c}TNO Science and Industry, Stieltjesweg 1, 2600 AD Delft, The Netherlands;
}

\authorinfo{Further author information: (Send correspondence to A.M.)\\A.M.: E-mail: alcione.mora@esa.int, Telephone: +34 91 813 1480}

\begin{document} 
  \maketitle 

\begin{abstract}
The ESA Gaia spacecraft has two Shack-Hartmann wavefront sensors (WFS) on its focal plane. They are required to refocus the telescope in-orbit due to launch settings and gravity release. They require bright stars to provide good signal to noise patterns. The centroiding precision achievable poses a limit on the minimum stellar brightness required and, ultimately, on the observing time required to reconstruct the wavefront. Maximum likelihood algorithms have been developed at the Gaia SOC. They provide optimum performance according to the Cram\'er-Rao lower bound. Detailed wavefront reconstruction procedures, dealing with partial telescope pupil sampling and partial microlens illumination have also been developed. In this work, a brief overview of the WFS and an in depth description of the centroiding and wavefront reconstruction algorithms is provided.
\end{abstract}




\section{Gaia, the need for refocusing}
\label{sect:gaiaRefocusing}

The ESA Gaia mission will provide astrometry of a billion objects in the Gaiaxy with unprecedent precision and accuracy. In addition, intermediate resolution spectra will be obtained for millions of sources. More details on the mission general goals can be found elsewhere\cite{LL:ESA-SCI(2000)4,2010SPIE.7731E..35D}.

\begin{figure}
\begin{center}
\includegraphics[height=5cm]{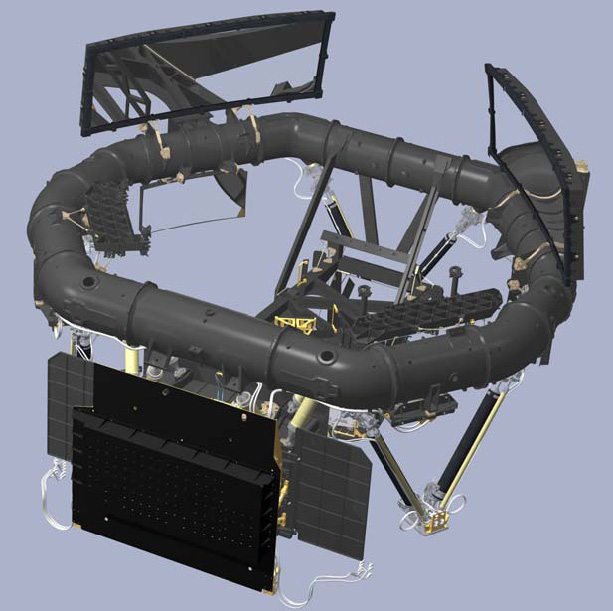}
\includegraphics[height=5cm]{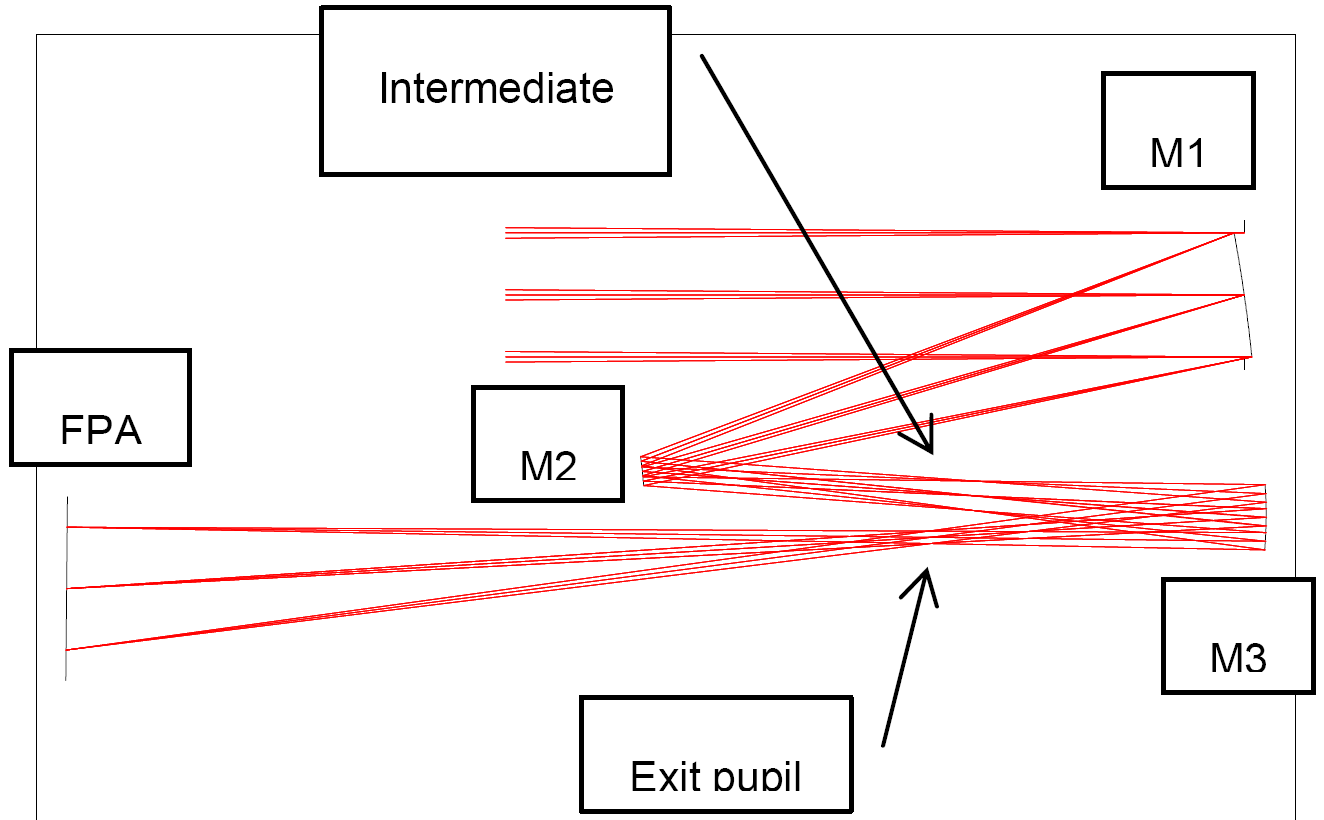}
\end{center}
\caption{Gaia payload overview (left) and unfolded telescope optical design (right). Courtesy Astrium.}
\label{fig:gaiaOptics}
\end{figure}

The payload is composed of two twin off-axis three mirror anastigmatic telescopes (TMA) with a rectangular pupil of 1.45$\times$0.5 m feeding a common focal plane\cite{2010SPIE.7731E..35D} (see also Kohley et al., this volume). In addition to the powered surfaces, several plane mirrors are required, two for the pupil plane beam combiner and two for a common periscope. Fig.~\ref{fig:gaiaOptics} provides an overview of the overall optical system. Most of the payload, including the mirrors, focal plane and torus support structure are made of silicon carbide (SiC). This material combines low weight with high stiffness and thermal conductivity, providing a very homogenous temperature in-orbit.

\begin{figure}
\begin{center}
\includegraphics[width=0.45\hsize]{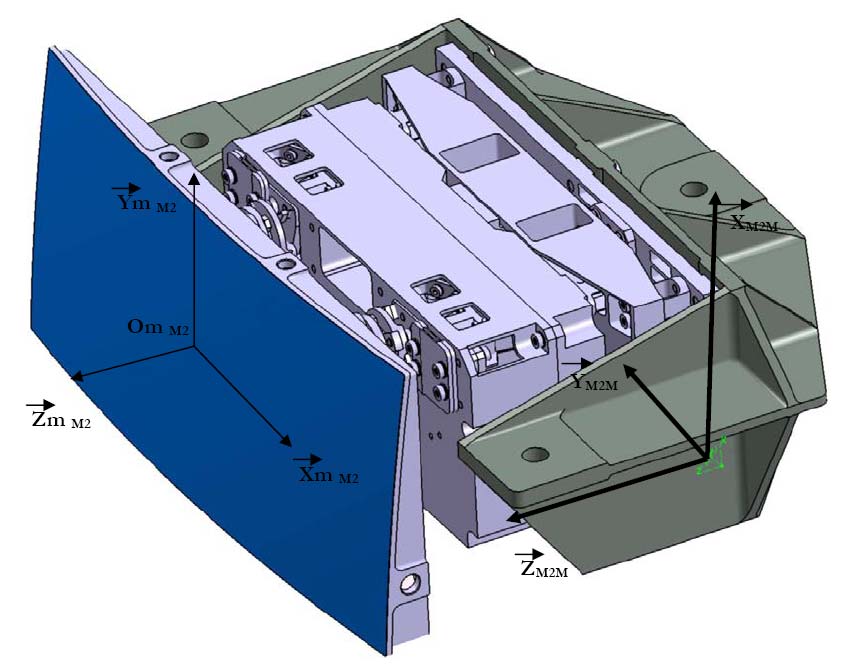}
\end{center}
\caption{M2 Movement Mechanism (M2MM). Courtesy Astrium.}
\label{fig:m2mm}
\end{figure}

Gaia will operate in the visible range (300-1050 nm) with a very high quality optical system (total wavefront error budget $\sim$50 nm). The mechanical tolerances for such a folded TMA system are tight, and smaller that the typical perturbations estimated for the launch vibrations and gravity release. Focusing mechanisms have thus been incoroporated to move each secondary mirror (M2), the so called M2 Movement Mechanisms (M2MM), built by SENER. Each M2MM has a fully redundant set of actuators capable of orienting the M2 surface with five degrees of freedom (which is enough for a rotationally symmetric surface). Fig.~\ref{fig:m2mm} shows a model of the system.

Detailed tolerancing simulations have been done at Astrium, with the M2MM as the sole compensator. They show that the design wavefront error can be recovered after actuation of the M2MM, with a very small residual contribution to the WFE (a few nm).

\section{The Gaia wavefront sensor}

\begin{figure}
\begin{center}
\includegraphics[width=0.75\hsize]{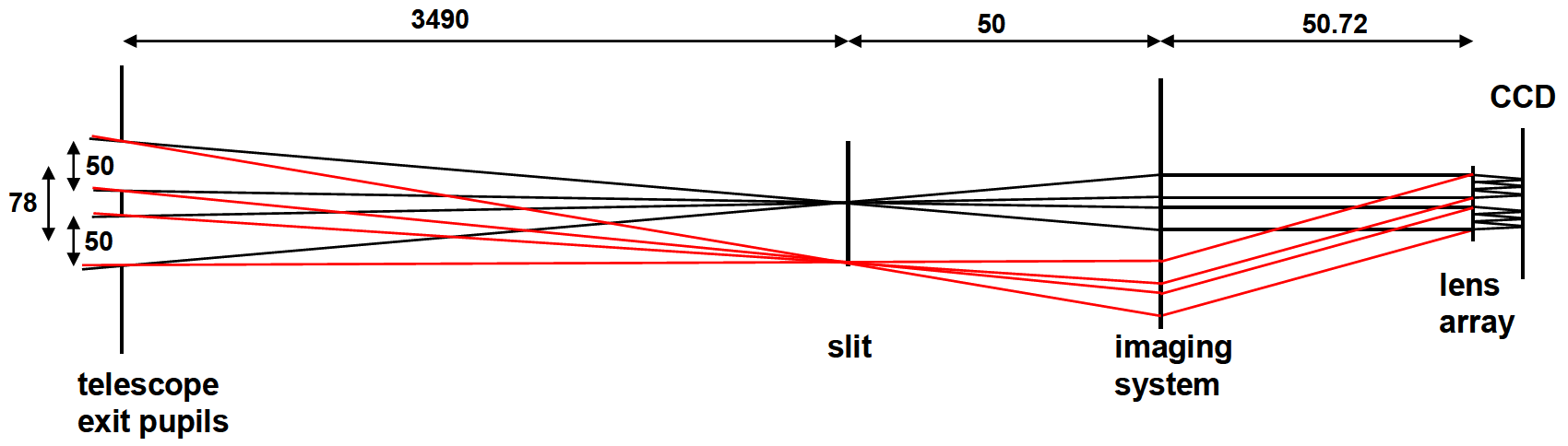}
\includegraphics[width=0.75\hsize]{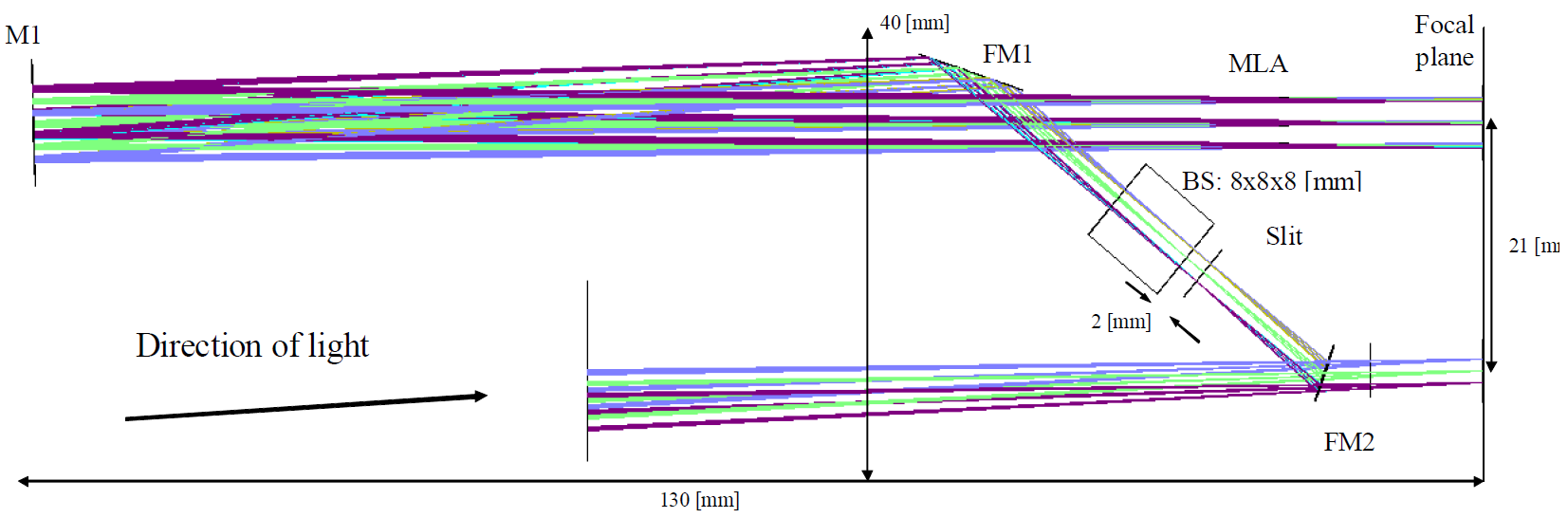}
\includegraphics[width=0.75\hsize]{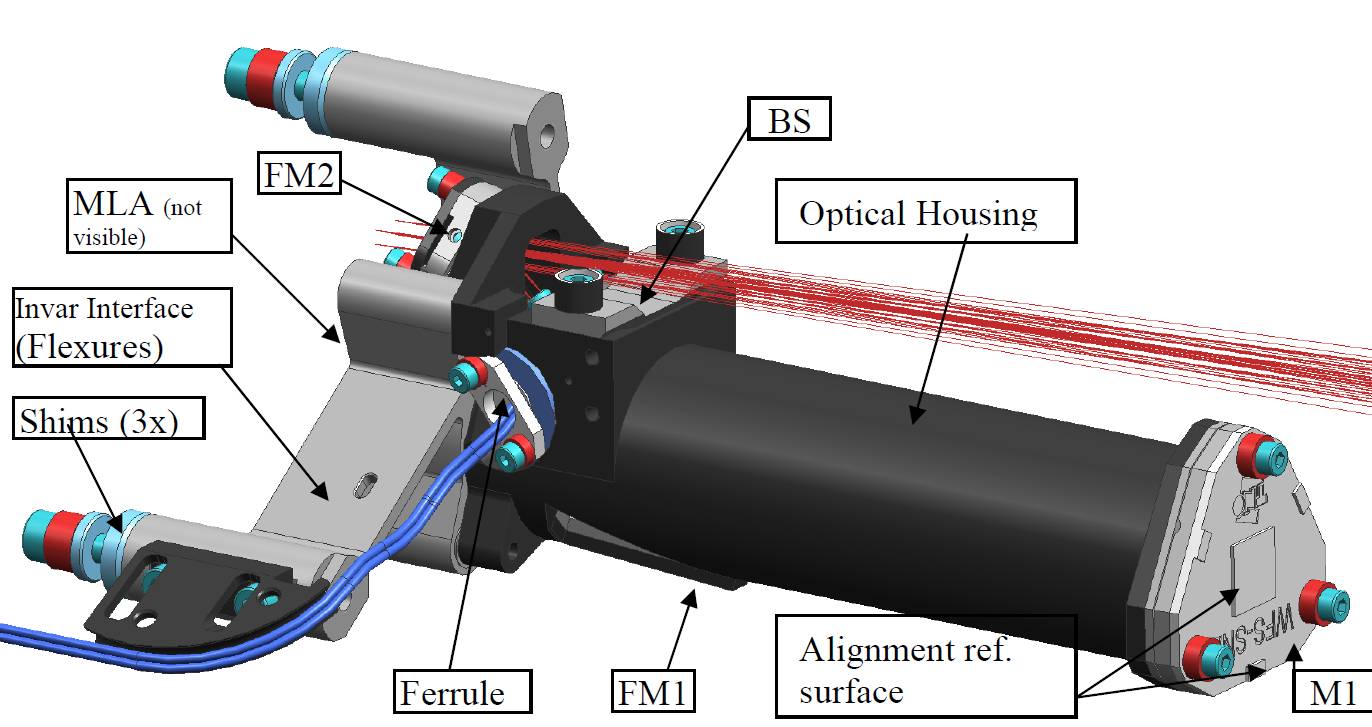}
\end{center}
\caption{Gaia wavefront sensor. Top: schematic layout of the WFS. From left to right, the gaia exit pupils and WFS entrance slit, collimator, microlens array and focal plane can be seen. Middle: optical design layout. The second fold mirror (FM2) deflects the incoming light to a beamsplitter (BS), where the reference signal from the optical fibres can be injected. The first fold mirror (FM1) sends the light to the collimator M1, which produces an image of the output pupil at the location of the microlens array (MLA). Finally, the lenslet images are projected onto the same focal plane used for the Gaia astrometric field. Bottom: mechanical overview of the WFS. Courtesy TNO\cite{2009SPIE.7439E..29V}.}
\label{fig:wfsOpticalDesign}
\end{figure}

Two Shack-Hartmann WaveFront Sensors (WFS), built by TNO, are located on the Gaia focal plane to provide the information required to drive the M2MM. The structure of the WFS is made of invar, while the optical surfaces are made from fused silica. Both materials provide a good thermal match to the SiC CCD support structure. The optical design of the WFS is displayed in Fig.~\ref{fig:wfsOpticalDesign}. It is based on an input slit (12''$\times$30''), an spherical collimator, a microlens array (387 $\mu$m pitch, 378 $\mu$m diameter), a beamsplitter cube and two fold mirrors. Each Gaia output pupil is sampled with an array of 3$\times$10 microlenses. The fold mirrors are required to project the WFS image plane at the location of the Gaia astrometric field image plane. That is, the WFS pattern is projected onto a CCD sharing the same focal plane than the other astrometric detectors (although mounted on reverse). The focal length of the microlenses is 18.5 mm, six times smaller than that for the collimator (111 mm), so the WFS CCD is operated with a TDI period six times larger. The zero wavefront error reference is provided by a set of three optical fibres, which mimic perfect point sources in the entrance slit thanks to the beamsplitter cube. They are fed by a SLED diode with 825 nm central wavelength. Additional information on the WFS is available in the literature\cite{2009SPIE.7439E..29V}.

\section{Centroids, Cram\'er-Rao and maximum likelihood}

The Shack-Hartmann wavefront sensor provides information on the wavefront slopes measuring the centroid displacement of each lenslet image with respect to a given zero error reference. This process critically depends on the centroiding precision achieavable. The higher the relative precision, the fainter the observed star can be. There is a limit on the maximum precision achievable for a centroiding algorithm: the Cram\'er-Rao lower bound. For a PSF or LSF observed with CCDs, it can be expressed as\cite{1978moas.coll..197L,LL:2004BASNOCODE,2010ISSIR...9..279L}

\begin{equation}
  \sigma_\eta = \frac{1}{\sqrt{ \displaystyle\sum_{k=0}^{n-1} \frac{ (S'_k)^2 }{r^2 + b + S_k} }}
\label{eq:cramerRaoLimit}
\end{equation}

where $\sigma_\eta$ is the Cram\'er-Rao centroiding uncertainty on the $\eta$ axis of an image, $S_k$ the discrete values of the PSF or LSF, in units of electrons, measured for each of the $k$ samples of the image (CCDs can be binned on-chip, and each sample can be larger or equal than one physical pixel), $S'_k$ is the derivative of $S_k$ with respect to the axis $\eta$ sample coordinate, $r$ the read-out noise (in electrons) and $b$ the homogeneous sky background (in electrons). The units of $\sigma_\eta$ are pixels. They can be converted to angles multiplying by the pixel size and dividing by the telescope focal length.

This equation has extensively been used to assess the performance achievable by Gaia. It is worth noting that only the largest contribution is from those samples for which the PSF or LSF slope is higher. The information contained in samples far from the PSF quickly goes to zero. Real or simulated PSFs can be used to estimate the Cram\'er-Rao centroiding precision, The derivatives can be estimated by nearest neighbours finite differences:

\begin{equation}
  S'_k \simeq \frac{ S_{k+1} - S_{k-1} }{2} 
\end{equation}

Reaching the Cram\'er-Rao precision is not trivial. It has been shown that a good way of doing it is using maximum likelihood estimators\cite{LL:LL-078}. In this approach, the sample data $\lbrace{S_k}\rbrace$ is fitted by a mathematical model $N_k(x_1, x_2, ..., x_n)$, which depends on several variables. One or two variables are centroid coordinates, for LSFs and PSFs, respectively. The other variables can be considered ``shape'' parameters (e.g. total flux, PSF width, microlens diameter, ...)  required to provide a good agreement between the model and the observations.

The problem of retrieving the centroid has thus been transformed in obtaining the set of parameters $\lbrace{x_1, x_2, ..., x_n}\rbrace$ providing the best fit. Maximum likelihood methods are those minimising the quantity

\begin{equation}
  \sum_k \frac{(S_k - N_k)^2}{r^2 + b + S_k}
\end{equation}

where $S_k - N_k$ is the difference between the actual value and the model for each sample $k$ and $\sqrt{r^2 + b + S_k}$ is the measurement error for $S_k$. This is just a weighted least squares minimisation problem. Many non-linear numerical methods are available to carry out such optimisation (e.g. Gauss-Newton, Levenberg-Marquardt). Some of them have been implemented in public domain libraries. In this work, the Java Apache Commons-Math\footnote{http://commons.apache.org/math/} implementation of the Levenberg-Marquardt algorithm has been used.

\section{Mathematical formulation \label{sect:maths}}

All least squares minimisation algorithms need an estimate of the derivatives of the PSF model with respect to each fitted variable:

\begin{equation}
  \frac{\partial N_k}{\partial x_i} =
  \frac{\partial N_k}{\partial x_i} (x_1, x_2, ..., x_n)
\end{equation}

Some algorithms approximate the derivatives applying finite differences on a sufficiently small interval. They are better suited for complex PSFs difficult to model. On the other hand, purely analytical models with an explicit expression for the derivatives, can be much faster. The applicability of one approach or the other depends on the complexity of the problem.

Well designed Shack-Hartmann sensors are typically diffraction limited. This means that the monochromatic PSF can be approximated by an Airy pattern. In the following, a collection of formulae will be given to estimate the number of electrons collected by a sample $N_{e^-}$ and its derivatives with respect to five parameters: the centroid coordinates $x$, $y$, the total flux $N$ and the lenslet diameter $D$ and focal length $f$. Both the monochromatic and polychromatic cases will be considered. The value of these formulae, whose derivation is straightforward, is to have all of them together to facilitate the implementation by an interested reader.

\subsection{Monochromatic PSF for a single microlens}
The monochromatic irradiance Airy pattern $I(x, y)$ produced by a diffraction limited lenslet is given by

\begin{equation}
  I(x, y, \lambda) = I(\rho) = I_0 \left({ \frac{2 J_1(\rho)}{ \rho } }\right)^2
                   = I_0 ( J_0(\rho)  + J_2(\rho) )^2
\end{equation}
\begin{equation}
  \rho = \frac { \pi D \sqrt{ (x - x_0)^2 + (y - y_0)^2 } }
               { \lambda f }
\end{equation}

where $J_n(r)$ is the $n$-th order Bessel function of the first kind, $\rho$ is the reduced radial coordinate, $x_0$ and $y_0$ the location of the PSF centre on the focal plane, $D$ the diameter of a microlens, $f$ the focal length of a microlens and $\lambda$ the wavelength.

\subsection{Monochromatic flux normalisation: number of electrons collected}

The monochromatic PSF formula given above is only valid for small angles. However, it quickly goes to zero and can thus be integrated over the whole focal plane to obtain the total number of electrons $N$

\begin{equation}
  N = \int_{-\infty}^\infty \int_{-\infty}^\infty dx dy I(x,y)
    = 2 \pi I_0 \left( \frac{\lambda f}{\pi D}\right )^2
      \int_0^\infty \rho d \rho \left( 2 \frac{ J_1(\rho) }{ \rho } \right)^2
    = \frac{4 I_0 \lambda^2 f^2}{\pi D^2}
\end{equation}

where the following orthogonality relation for Bessel functions has been used

\begin{equation}
  \int_0^\infty J_\alpha(x) J_\beta(x) \frac{dx}{x}
  = \frac{2}{\pi} \frac{ \sin \frac {\pi(\alpha - \beta)} {2} }{ \alpha^2 - \beta^2 }
\end{equation}

and the units of $I_0$ are electrons per unit area

\begin{equation}
  I_0 = \frac{ N \pi D^2 }{ 4 \lambda^2 f^2 }
\end{equation}

\subsection{Derivatives with respect to monochromatic PSF parameters}

The derivatives of the intensity $I(x, y, \lambda)$ with respect to the five variables $x_0$, $y_0$, $D$, $f$ and $I_0$ that define the location, width and intensity are given below. Properties of the Bessel functions have been used to remove $\rho$ from the denominator and avoid singularities near the origin. 

\begin{equation}
  \begin{array}{ll}
    \displaystyle \frac{\partial I}{\partial x_0} (x, y, \lambda)
      & \displaystyle = - I_0 ( J_0(\rho) + J_2(\rho) ) ( J_1(\rho) + J_3(\rho) ) 
        \frac{ \pi^2 D^2 (x_0 - x) }{ \lambda^2 f^2 \rho } \\
      \displaystyle
      & \displaystyle = - I_0 ( J_0(\rho) + J_2(\rho) )
        ( 3 J_0(\rho) + 4 J_2(\rho) + J_4(\rho) ) 
        \frac{ \pi^2 D^2 (x_0 - x) }{ 6 \lambda^2 f^2 }
  \end{array}
\end{equation}
\begin{equation}
  \begin{array}{ll}
    \displaystyle \frac{\partial I}{\partial y_0} (x, y, \lambda)
      & \displaystyle = - I_0 ( J_0(\rho) + J_2(\rho) ) ( J_1(\rho) + J_3(\rho) ) 
        \frac{ \pi^2 D^2 (y_0 - y) }{ \lambda^2 f^2 \rho } \\
      \displaystyle
      & \displaystyle = - I_0 ( J_0(\rho) + J_2(\rho) )
        ( 3 J_0(\rho) + 4 J_2(\rho) + J_4(\rho) ) 
        \frac{ \pi^2 D^2 (y_0 - y) }{ 6 \lambda^2 f^2 }
  \end{array}
\end{equation}
\begin{equation}
  \frac{\partial I}{\partial D} (x, y, \lambda)
    = - I_0 ( J_0(\rho) + J_2(\rho) ) ( J_1(\rho) + J_3(\rho) )
      \frac{ \rho }{ D }
\end{equation}
\begin{equation}
  \frac{\partial I}{\partial f} (x, y, \lambda)
    = - I_0 ( J_0(\rho) + J_2(\rho) ) ( J_1(\rho) + J_3(\rho) ) 
      \frac{ -\rho }{ f }
\end{equation}
\begin{equation}
  \frac{\partial I}{\partial I_0} (x, y, \lambda)
    = ( J_0(\rho)  + J_2(\rho) )^2
\end{equation}

\subsection{Non-SI units}

The following non-SI units have been used to make the interpretation of the results easier: The total number of electrons collected $N$ instead of the peak intensity $I_0$, the lengths are measured in samples ($x_{\rm sample}$, $y_{\rm sample}$), instead of metres ($x$, $y$). The derivatives with respect to these derived magnitudes can be computed using the chain rule.

\begin{equation}
  \frac{ \partial I }{ \partial N }
    = \frac{ \pi D^2 }{ 4 \lambda^2 f^2 } \frac{\partial I}{\partial I_0}
\end{equation}
\begin{equation}
  \frac{ \partial I }{ \partial x_{\rm sample} }
    = \frac{ \partial I }{ \partial x } \Delta x_{\rm }
\end{equation}
\begin{equation}
  \frac{ \partial I }{ \partial y_{\rm sample} }
    = \frac{ \partial I }{ \partial y } \Delta y_{\rm }
\end{equation}

\subsection{Image composition and number of electrons per sample}

The monochromatic images including the contribution for all the lenslets in the WFS will be modeled as the addition of $n_{\rm pupil} = 3 \times 10$ components per telescope and a diffuse light homogeneous background. Partially illuminated lenslets have not been considered, because they are difficult to model.

\begin{equation}
  I_{\rm total}(x, y, \lambda)
    = \sum_{i=0}^{n_{\rm pupil} - 1} {I_i(x, y, \lambda) + {\rm sky}}
\end{equation}

\subsection{Number of electrons per sample: integration}

The monochromatic number of electrons $N_{e^-}(i, j, \lambda)$ collected at the sample $(i,j)$ and wavelength $\lambda$, and its derivatives $\frac{ \partial N_{e^-}(i, j, \lambda) }{ \partial \xi }$ with respect to any variable $\xi_i$ are computed integrating on the pixel rectangular boundaries:

\begin{equation}
  N_{e^-}(i, j, \lambda)
    = \int_{i \Delta x_{\rm }}^{(i+1) \Delta x_{\rm }} dx
      \int_{j \Delta y_{\rm }}^{(j+1) \Delta y_{\rm }} dy
      \left({ \sum_{i=0}^{n_{\rm pupil} - 1} {I_i(x, y, \lambda) + {\rm sky}} }\right)
\end{equation}
\begin{equation}
  \frac{ \partial N_{e^-}(i, j, \lambda) }{ \partial \xi_i }
    = \int_{i \Delta x_{\rm }}^{(i+1) \Delta x_{\rm }} dx
      \int_{j \Delta y_{\rm }}^{(j+1) \Delta y_{\rm }} dy
      \frac{ \partial I(x, y, \lambda) }{ \partial \xi_i }
\end{equation}

where

\begin{equation}
  \frac{ \partial I(x, y, \lambda) }{ \partial \xi_i }
    = \left\{{
        \begin{array}{ll}
          \displaystyle\frac{ \partial I_i(x, y, \lambda) }{ \partial \xi_i }
            &; \xi_i {\rm \;is\;a\;variable\;for\;lenslet\;component\;} i\\
          \\
          \displaystyle\frac{1}{ \Delta x_{\rm } \Delta y_{\rm } }
            &; \xi_i \equiv {\rm sky}
        \end{array}
     }\right.
\end{equation}

\subsection{Polychromatic calculation}

The expressions above are valid for monochromatic or quasi-monochromatic light sources, such as the SLEDs feeding the WFS calibration optical fibres. However, an additional integration in wavelength is needed to produce a WFS stellar image.

\begin{equation}
  N_{e^-}(i, j) = \int_0^\infty N_{e^-}(i, j, \lambda) w(\lambda) d\lambda
\end{equation}
\begin{equation}
  w(\lambda)
          = \frac {F (\lambda) T (\lambda)}
                     {\int_0^\infty F (\lambda) T (\lambda) d\lambda}
     \simeq \frac {F (\lambda) R^9 (\lambda) {\rm QE} (\lambda)}
                     {\int_0^\infty F (\lambda) R^9 (\lambda)
                       {\rm QE} (\lambda) d\lambda}
\end{equation}

where $w(\lambda)$ is the weighting function, $F (\lambda)$ is the stellar spectrum, in units of photons per unit area per unit time, $T (\lambda)$, is the Gaia + WFS spectral response funtion, $R (\lambda)$ the reflectivity of a single protected silver mirror, 9 the number of mirrors in the telescope-WFS optical path (6 for the telescope and 3 for the WFS) and QE($\lambda$) the CCD quantum efficiency. The wavelength dependence of the MLA overall transmission has been neglected. The integration in wavelength has been approximated by a sum over discrete wavelengths

\begin{equation}
  N_{e^-}(i, j) \simeq
    \sum_{k = 0}^{n_\lambda - 1} N_{e^-}(i, j, \left\langle{\lambda}\right\rangle_k) w_k
\end{equation}
\begin{equation}
  w_k = \frac{ \left\langle{ F(\lambda)        }\right\rangle
                \left\langle{ R(\lambda)        }\right\rangle^6
                \left\langle{ {\rm QE}(\lambda) }\right\rangle }
              {\displaystyle\sum_{l = 0}^{n_\lambda - 1} w_l}
\end{equation}

where the averages are carried out in $n_\lambda$ intervals of equal length.

\section{Simulations: image generation and centroiding precision
\label{sect:simulationsCentroid}}

An image simulator and analyser has been developed at the Gaia SOC to produce artificial WFS images and test the fitting algorithm presented. It has been coded in Java, using existing libraries, such as those already developed by the Gaia Data Processing and Analysis Consortium (DPAC)\cite{LL:FM-030}. The least squares optimiser has been taken from Apache Commons Math, while the Bessel functions are computed with the Colt library\footnote{http://acs.lbl.gov/software/colt/}.

Pixel integrations use a single interval Gauss-Legendre algorithm with 3 AL $\times$ 4 AC knots per pixel sampling. Monochromatic and polychromatic calculations are possible. The former are appropriate for the calibration sources, while the latter can be used for stellar sources. The polychromatic calculations cover the wavelength range 300-1050 nm, which has been divided into 10 sub-intervals. A reference G2V spectrum has been used to simulate stellar sources. Poisson shot noise and Gaussian read-out noise can be added to the perfect noiseless images to simulate real observations.

\begin{figure}
\begin{center}
\includegraphics[width=0.75\hsize]{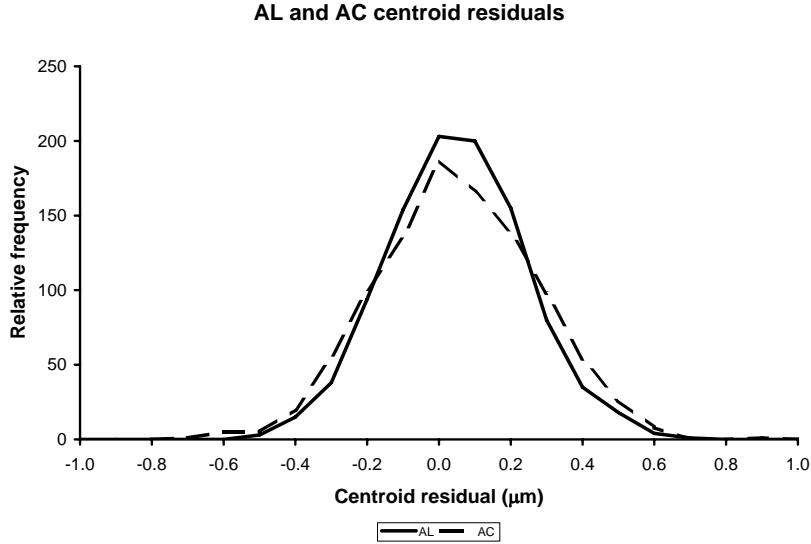}
\end{center}
\caption{Centroid residuals. Single image Gaia WFS centroid fitting resituals are shown. The performance is only slightly better along scan, in spite of the pixel size being three times smaller. This shows that maximum likelihood centroiding can provide good results even with slightly undersampled images. The results agree withing 5\% with the Cram\'er-Rao estimations carried out on the images according to Eq.~\ref{eq:cramerRaoLimit}. This demonstrates the method used is, indeed, optimal.}
\label{fig:centroidResiduals}
\end{figure}

A set of 1000 monochromatic PSFs have been simulated with different centroid locations and noise. The total number of electrons is 10000, similar to the number of electrons needed to provide a good sampling of the gaia wavefront error for a perfect algorithm. The central SLED wavelength of {830}{nm} has been used. They have subsequently been fitted by the maximum likelihood centroiding algorithm. The fit residuals are shown in Fig.~\ref{fig:centroidResiduals}. Because of the rectangular pixel size, they have been converted to $\mu$m to provide a common comparison scale.

It can be appreciated that the PSF sampling only has a minor impact: the AL centroiding precision is only $\sim$16\% better, despite the pixel size being three times smaller. The centroid uncertainties found agree very well (within 5\%) with the estimates provided by the application of Eq.~\ref{eq:cramerRaoLimit} to the observed PSFs. The values obtained, however, are worse ($\sim$47\% AL, $\sim$70\% AC)  than those obtained applying the following equation

\begin{equation}
  \sigma_x \simeq \frac{\lambda f}{\sqrt{N} \pi D} = 0.129 \mu \rm m
\end{equation}

which is the limit for a PSF sampled with infinitesimally small pixels and negligible read-out noise \cite{1978moas.coll..197L,LL:2004BASNOCODE}. It is worth noting that this expression provides a 30\% better precision than that obtained for a gaussian function with a width equal to the Airy disc core. That is, significant centroiding information is stored in the Airy disc secondary maxima.

Three conclusions can be drawn from this experiment. First, the centroiding method used is indeed optimal, and no further precison can be gained using more sophisticated algorithms. Second, simple rules of thumb for well sampled pure Poisson noise image models are not directly applicable to obtain the centroiding precision achievable under conditions of PSF undersampling or significcant read-out noise. Third, accurate predictions on the performance of a real wavefront sensor can be obtained computing the Cram\'er-Rao limit according to Eq.~\ref{eq:cramerRaoLimit} on simulated images.

In addition to simple PSF modelisation, full telescope pupil mono- and polychromatic simulations have been run to demonstrate the applicability to scenarios similar to real operations. The full WFS combined image is the sum of the Airy patterns corresponding to each lenslet.

\begin{figure}
\begin{center}
\includegraphics[width=\hsize]{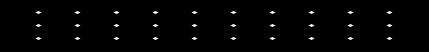}
\end{center}
\caption{WFS full pupil pattern. $3 \times 10$ sub-pupil polychromatic images are located in a perfect rectangular network, with a 39 pix AL and 13 pix AC unit cell.
\label{fig:wfsPattern}}
\end{figure}

For a $3 \times 10$ sub-pupils image (partially illuminated lenslets are not considered), a total of 121 variables have to simultaneously be fitted by the algorithm. Two tests (monochromatic and polychromatic) have been carried out to assess if this is, in fact, possible. The lenslet images are arranged in a perfect $3 \times 10$ network, with a 39 pix AL and 13 pix AC unit cell. The total signal is 10000 electrons per lenslet. Fig.~\ref{fig:wfsPattern} shows an illustrative WFS synthetic image. The polychromatic calculations have used the G2V stellar template.

Both the mono- and polychromatic calculations converge. The time required for this brute-force optimisation is 8.8 min and  80 min, respectively for a regular desktop Core2 Duo at 3 GHz. However, the computation speed can significantly be increased applying the following recipes.

\begin{enumerate}
\item Fitting each lenslet image independently. This is a reasonable assumption for most wavefront sensors and, in particular, for Gaia, where the overlap between neighbour subimages is below 1 $e^-$. It can be done either solving 3$\times$10 different problems or limiting the extension of the PSF and its derivatives.
\item Using Look Up Tables (LUTs) to store values for the PSF values and its derivatives. Interpolating inside LUTs is much faster than Bessel function evaluation and integration, and can be generalised to different PSF models.
\item Using modified least squares algorithms where LUTs are also used to store the matrix used to estimate the parameter update for each iteration.
\end{enumerate}

All the suggestions above have been implemented for the Gaia WFS. The times required to fit a pattern have thus been reduced to around a minute for a single pattern applying suggestions 1 and 2 and down to a ms for a single image applying all the optinisations explained. It also seems that using the massive paraellel computing power of modern FPGAs or GPUs, maximum likelihood algorithms could be applied to the much more demanding task of real time closed loop adaptive optics wavefront sensing.

\section{WFS wavefront reconstruction}

Typical commercial Shack-Hartmann wavefront sensors for on-ground applications have hundreds or thousands of lenslets. When these devices, providing a dense pupil sampling, are carefully aligned to the pupil and focal plane axes, the wavefront can easily be reconstructed. In fact, manufacturers usually provide ad-hoc custom software providing either the raw wavefront or its expansion into a polynomial series.

The case for Gaia, however, is more complicated. On the one hand, only a small number of lenslets (3$\times$10 fully illuminated lenslets) per pupil is available. By design, the lenslet pitch was chosen to be a quarter of the pupil short dimension. This guarantees that 3$\times$10 elements are always fully illuminated, although they will not cover the full pupil. This choice makes the device much less sensitive to microlens alignment errors or launch settings, at the price of leaving a significant portion of the telescope pupil unsampled. In addition, the Gaia WFS microlens array will only be rougly aligned with the telescope output pupil and the CCD.

The objective for the Gaia in-flight WFSs is to measure low order aberrations due to launch setting and gravity realease, so a sparse pupil sampling is enough. However, some usual simplifying assumptions are now not possible. In particular, the real location of each lenslet with respect to the telescope pupil and focal plane must be determined and considered. In addition, each lenslet subtends a significant portion of the output pupil, and they may not be considered infinitesimally small.

A custom wavefront reconstruction algorithm has been developed. It takes many ideas and definitions from previous work by Astrium. The wavefront is expressed as a series of bidimensional Legendre polynomials, the first of which are given in Table~\ref{tab:legendre}. In addition, it directly converts the slopes into the polynomial expansion without reconstructing the wavefront as an intermediate step. The microlenses are not considered point-like. Pupil integration is carried out in a sort of ray tracing matrix computation. This method could be useful to other missions with few elements WFS.

\begin{table}
\caption{Bidimensional Legendre polynomials. The WFS decomposes the WFE into a series of Legendre polynomials. The first nine terms are given below. The first three have no impact on image quality (piston, tilt $x$ and tilt $y$).}
\label{tab:legendre}
\begin{center}
\begin{tabular}{lll}
\hline
\hline
Polynomial  & 1-D product     & Formula \\
\hline
$L_1 (x,y)$ & $P_0(x) P_0(y)$ & 1                          \\
$L_2 (x,y)$ & $P_0(x) P_1(y)$ & $x$                        \\
$L_3 (x,y)$ & $P_1(x) P_0(y)$ & $y$                        \\
$L_4 (x,y)$ & $P_2(x) P_0(y)$ & $\frac{1}{2} (3x^2 - 1)$   \\
$L_5 (x,y)$ & $P_1(x) P_1(y)$ & $xy$                       \\
$L_6 (x,y)$ & $P_0(x) P_2(y)$ & $\frac{1}{2} (3y^2 - 1)$   \\
$L_7 (x,y)$ & $P_0(x) P_3(y)$ & $\frac{1}{2} (5y^2 - 3y)$  \\
$L_8 (x,y)$ & $P_2(x) P_1(y)$ & $\frac{1}{2} (3x^2 - 1) y$ \\
$L_9 (x,y)$ & $P_1(x) P_2(y)$ & $\frac{1}{2} x (3y^2 - 1)$ \\
\hline
\end{tabular}
\end{center}
\end{table}

\section{Wavefront Legendre decomposition}

Three reference systems must be defined. First, the telescope pupil reference system $\mathcal{R}_{\rm pupil} (O_{\rm pupil}, x, y)$, where $O_{\rm pupil}$ is located in the centre of the WFS pupil, $x$ and $y$ are parallel to the exit pupil minor and major axes, respectively. and are normalised coordinates: $x, y \in [-1, 1]$. Second, the lenslet reference system $\mathcal{R}_{\rm lenslet} (O_{\rm lenslet}, x', y')$ has axes parallel to the directions upon which the lenslets are distributed. The $x'$ and $y'$ directions are approximately parallel to the WFS microlens array minor and major axes $x$, $y$, respectively. Third, the CCD reference system $\mathcal{R}_{\rm CCD} (O_{\rm lenslet}, x'', y'')$, where the centroids are estimated. The transformation relations between them are as follows

The transformation between $\mathcal{R}_{\rm pupil}$ and $\mathcal{R}_{\rm lenslet}$ is
\begin{equation}
  \left(\begin{array}{l} x \\ y \end{array}\right)
  = \left(\begin{array}{rr}
        2 \cos\theta / D_x & 2 \sin\theta / D_x \\
       -2 \sin\theta / D_y & 2 \cos\theta / D_y
    \end{array}\right)
    \left(\begin{array}{l} x' \\ y' \end{array}\right)
  + \left(\begin{array}{l} \Delta x \\ \Delta y \end{array}\right)
\end{equation}

where $\theta$ is the angle between the lenslet array and the WFS pupil, $D_x$ and $D_y$ are the WFS pupil dimensions and $(\Delta x, \Delta y)$ are the offsets between the centre of the microlens array and the WFS pupil in normalised coordinates. The transformation between $\mathcal{R}_{\rm lenslet}$ and $\mathcal{R}_{\rm CCD}$ is
\begin{equation}
  \left(\begin{array}{l} x' \\ y' \end{array}\right)
  = \left(\begin{array}{rr}
        \cos\theta' & \sin\theta' \\
       -\sin\theta' & \cos\theta'
    \end{array}\right)
    \left(\begin{array}{l} x'' \\ y'' \end{array}\right)
\end{equation}

where $\theta'$ is the rotation angle between the microlens array and the CCD.

For a given field of view, the wavefront error (WFE) is a bidimensional function of the WFS normalised pupil coordinates $(x, y)$.
\begin{equation}
  {\rm WFE} = {\rm WFE}(x, y) \sum_{j = 1}^{n_L} \alpha_j L_j (x, y)
\end{equation}
\begin{equation}
  j = 1, 2, ..., n_L
\end{equation}

where the wavefront is expressed as a series of bidimensional Legendre polynomials. $n_L = 9$ is the number of 2-D Legendre polynomials used. Legendre polynomials are particularly convenient because they are orthogonal in the range [-1,1]. This means that the RMS WFE of the Legendre expansion is:
\begin{equation}
  {\rm RMS}
    = \sqrt{ \left\langle{\rm WFE^2}\right\rangle - \left\langle{\rm WFE}\right\rangle^2 }
    = \sqrt{ \sum_{\scriptsize \begin{array}{c} j = 1 \\ d_{xj} + d_{yj} > 1 \end{array}}^{n_L}
        \frac{\alpha_j^2}{ (2d_{xj} + 1) (2d_{yj} + 1) } }
\end{equation}

where $d_{xj}$ and $d_{yj}$ are the unidimensional Legendre orders in $x$ and $y$ and the sum is extended to all terms other than piston and tilt: $d_{xj} + d_{yj} > 1$.

The WFS samples the wavefront surface at $n$ locations in the exit pupil, corresponding to each  lenslet. In fact, the WFS measures, for each lenslet $i = 1, 2, ... n$, the centroid location differences $(\delta x''_i, \delta y''_i)$  between the images produced by the aberrated telescope and a reference zero WFE source (calibration optical fibre). For a lenslet $i$, the centroid displacement $(\delta x_i, \delta y_i)$ is the average of the wavefront error slope over the lenslet pupil area multiplied by the lenslet focal length $f$:
\begin{equation}
  \delta x'_i = \frac{f}{A_{\rm lenslet}}
               \int_{{\rm lenslet} \: i} dx' dy' \frac {\partial {\rm WFE} (x', y') } {\partial x'}
\end{equation}
\begin{equation}
  \delta y'_i = \frac{f}{A_{\rm lenslet}}
               \int_{{\rm lenslet} \: i} dx' dy' \frac {\partial {\rm WFE} (x', y') } {\partial y'}
\end{equation}
\begin{equation}
  \left(\begin{array}{l} \delta x' \\ \delta y' \end{array}\right)
  = \left(\begin{array}{rr}
        2f \cos\theta / D_x & -2f \sin\theta / D_y \\
        2f \sin\theta / D_x &  2f \cos\theta / D_y
    \end{array}\right)
    \left(\begin{array}{l}
      \displaystyle \frac{1}{ A_{\rm lenslet}}
        \int_{{\rm lenslet} \: i} dx' dy' \frac {\partial {\rm WFE} (x, y) } {\partial x} \\
      \displaystyle \frac{1}{ A_{\rm lenslet}}
        \int_{{\rm lenslet} \: i} dx' dy' \frac {\partial {\rm WFE} (x, y) } {\partial y}
    \end{array}\right)
\label{eq:legendreDecomposition}
\end{equation}

where the angular brackets in Eq.~\ref{eq:legendreDecomposition} denote average over a given subpupil of the the derivative of a given 2D Legendre polynomial. The number of points used in the average is equivalent to a number of rays ``traced'' for each polynomial. The lenslet integrals can reformulated as
\begin{equation}
  \left(\begin{array}{l} \delta x' \\ \delta y' \end{array}\right)
  = \left(\begin{array}{rr}
        2f \cos\theta / D_x & -2f \sin\theta / D_y \\
        2f \sin\theta / D_x &  2f \cos\theta / D_y
    \end{array}\right)
    \left(\begin{array}{l}
      \displaystyle \left\langle{ \frac {\partial {\rm WFE} (x, y) } {\partial x} }\right\rangle \\
      \displaystyle \left\langle{ \frac {\partial {\rm WFE} (x, y) } {\partial y} }\right\rangle
    \end{array}\right)
\end{equation}
\begin{equation}
  \left(\begin{array}{l} \delta x' \\ \delta y' \end{array}\right)  
  = \sum_{j = 1}^{n_L} \alpha_j
    \left(\begin{array}{rr}
        2f \cos\theta / D_x & -2f \sin\theta / D_y \\
        2f \sin\theta / D_x &  2f \cos\theta / D_y
    \end{array}\right)
    \left(\begin{array}{l}
      \displaystyle \left\langle{ \frac {\partial {L_j (x, y) } }{\partial x} }\right\rangle \\
      \displaystyle \left\langle{ \frac {\partial {L_j (x, y) } }{\partial y} }\right\rangle
    \end{array}\right)
\end{equation}

and can be expressed in matrix form
\begin{equation}
  \vec{\delta} = \matrix{B} \vec{\alpha}
\end{equation}
\begin{equation}
  \vec{\alpha} = \matrix{B}^+ \vec{\delta}
\end{equation}

where $\vec{\alpha}$ is a vector with the $\lbrace{ \alpha_j }\rbrace$ coefficients of the Legendre expansion, that is the final output of the wavefront reconstruction process, that forms an overdetermined system of linear equations together with the matrix $\matrix{B}$ and the centroid displacements vector $\vec{\delta}$. Solving that system of equations is equivalent to reconstructing the wavefront. $\matrix{B}^+$ is the pseudoinverse matrix and can be computed using e.g. singular value decomposition (SVD) techniques.

This method has the advantage that it could, in principle, be applied to partially illuminated lenslets just clipping the rays outside the telescope pupil from the average over the 2D Legendre polynomials. There are several ways of sampling an optical pupil with discrete rays. The most popular ones are the rectangular grid (default scheme in Code V) and the Gaussian quadrature polar grid (default scheme in Zemax). However, none of these methods provide an equal areas sampling scheme free of border effects for a circular pupil. The custom approach used in this work is described in the following.

The unit circle is divided into annuli. Each of them is divided into equal area annular sectors. The radial and angular dimension of each sector is made as similar as possible. For each annular sector, the ray passing through the center of mass is used. For a given number of annuli $n_R$ in the unit circle, with approximately equal radial extension, the number of sectors $n_i$ for the annulus $i$ that makes the radial and angular dimensions most similar can be estimated as:
\begin{equation}
  n_i = {\rm round}( 2 \pi i )
\end{equation}

The area $A_i$ for a sector in annulus $i$, with inner and outer radii $r_{i-1}$, $r_i$ is
\begin{equation}
  A_i = \frac{\pi(r_i^2 - r_{i-1}^2)}{n_i}
\end{equation}

If the area for all sectors is equal, it can be shown that
\begin{equation}
  r_i^2 = 
  \frac {\displaystyle\sum_{j = 0}^i n_j} {\displaystyle\sum_{j = 0}^{n_R} n_j}
\end{equation}

Finally, the radial center of mass $r_{\rm cm}$ of an annular sector limited by $r_{i-1}$ and $r_i$ in an annulus with $n_i$ sectors is given by
\begin{equation}
  r_{\rm cm}
    = \frac{\displaystyle \int_{-\pi/n_i}^{\pi/n_i} d\theta \int_{r_{i-1}}^{r_i} r dr r \cos \theta }
           {\displaystyle\frac{\pi}{n_i}(r_i^2 - r_{i-1}^2)}
    = \frac{2 n_i \sin\displaystyle\frac{\pi}{n_i}}
           {3 \pi} \frac{r_i^3 - r_{i-1}^3}{r_i^2 - r_{i-1}^2}
\end{equation}

\begin{figure}
\begin{center}
\includegraphics[width=0.75\hsize]{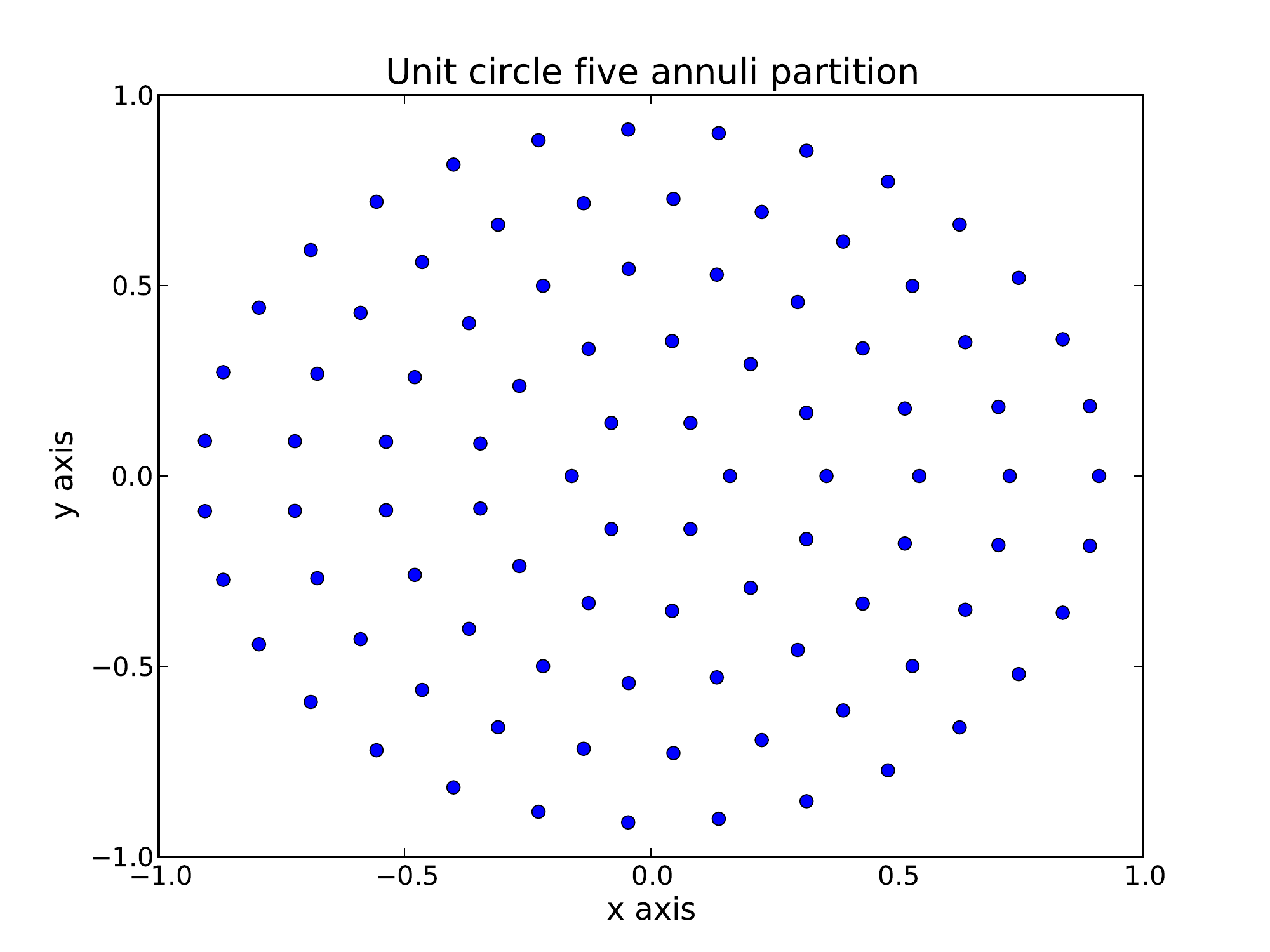}
\end{center}
\caption{Circular pupil partition. Five annuli have been used. The centre of mass of the 94 annular sectors covering the unit circle is shown. They provide a quite homogeneous sampling. Rays outside the telescope exit pupil are not clipped in the wavefront reconstruction process. This is the default sampling scheme used in this work.}
\label{fig:pupilPartition}
\end{figure}

Fig.~\ref{fig:pupilPartition} shows an example of circular pupil partition using five annuli. The centre of mass of the 94 annular sectors is shown on the unit circle. It is clear that the sampling is quite homogeneous.
 
\section{Simulations: wavefront reconstruction}

Wavefront generation and fitting capabilities have been added to the Java package developed for Gaia WFS data analysis following the approach described before. The method has been tested in two different scenarios.
\begin{enumerate}
\item The first one uses a wavefront composed of just the nine Legendre polynomial terms shown in Table~\ref{tab:legendre}.
\item The second one is similar to the first one, but adding a very large RMS WFE, to simulate the (mostly) low frequency post-launch effects.
\end{enumerate}

For the first set of simulations 1000 different wavefront 2D Legendre expansions have been generated. Each one is the sum of the first six non-trivial Legendre terms shown in Table~\ref{tab:legendre}. Each term is scaled by a random gaussian factor. The total WFE has been set to the whole low frequency error budget of $\sim$43.9 nm. Noiseless microlens image centroid displacements have been calculated and fitted using the algorithm presented before. The results obtained (see Table~\ref{tab:wavefrontReconstruction}) are very good, and the reconstruction error is negligible.

The second simulations mimic the post-launch misalignment using the same approach above, but increasing the RMS WFE to a pessimistic value of 1 $\mu$m. The results are also shown in Table~\ref{tab:wavefrontReconstruction}. It is clear that the Gaia WFS is capable of correctly identifying and measuring these aberrations. The telescopes could, in principle, be realigned using a single day of WFS observations.

\begin{table}[b]
\caption{Legendre fitting average residual errors for two different scenarios. 1000 different simulations have been run for each case. First, only low frequency terms compatible with the Gaia error budget. Second, large low frequency aberrations (up to 1 $\mu$m) simulating bad post-launch condition. The units are nanometers. No noise has been introduced in the simulations. It is apparent that the wavefront reconstruction works flawlessly for low frequency aberrations.
}
\label{tab:wavefrontReconstruction}
\begin{center}
\small
\begin{tabular}{lllllll}
\hline
\hline
Method           & $L_4$    & $L_5$    & $L_6$    & $L_7$    & $L_8$    & $L_9$    \\
\hline
1. Low frequency & 6.25e-14 & 5.47e-14 & 2.14e-14 & 3.29e-14 & 7.99e-14 & 6.33e-14 \\
2. After launch  & 1.44e-12 & 1.23e-12 & 4.86e-13 & 7.55e-13 & 1.78e-12 & 1.42e-12 \\
\hline
\end{tabular}
\end{center}
\end{table}

\section{Conclusions
\label{sect:conclusions}}

A maximum likelihood algorithm has been developed to determine the centroid of the sub-pupil images in the Gaia Shack-Hartmann wavefront sensors. Each subpupil is assumed to produce a perfect Airy diffraction patterns. The formulation for both the mono- and polychromatic cases is presented. The centroid location precision achived reaches the maximum performance possible dictated by the Cram\'er-Rao lower bound. Full pupil $3 \times 10$ images have been successfully fitted, demonstrating the applicability to real-case scenarios. Further optimization is possible separating the problem for each lenslets and using look-up tables bot for the PSF and intermediate matrices. Millisecond performance has been demonstrated for a single lenslet on a desktop computer.

A wavefront reconstruction algorithm for wavefront sensors with few lenslets not sampling whole rectangular apertures have been developed. The location of each lenslet in the telescope pupil is considered. Ray tracing integration over each leanslet is carried out to account for the significant size of each element as compared to the telescope output pupil. The method has successfully been applied to the Gaia case, where it has been demonstrated that large post-launch low order aberrations can easily be identified and expressed as a series of bidimensional Legendre polynomials. Finally, the algorithms presented will be applied to real WFS data in the near future.

\section{Acknowledgements}

The authors wish to thank Astrium and TNO for their support and access to internal documents on the wavefront sensor and Gaia optical design. Some concepts and ideas presented here come from those sources. The authros would also like to thank Matthias Erdmann and Ralf Kohley at ESA for their support, advice and fruiful discussions.

\bibliography{2012_06_spie_wfs,gaia_livelink_valid,gaia_livelink_obsolete,gaia_drafts,gaia_refs,gaia_books,gaia_refs_ads}   

\begin{thebibliography}{1}

\bibitem{LL:ESA-SCI(2000)4}
{GAIA~Science~Advisory~Group}, ``{G}{A}{I}{A}. {C}omposition, {F}ormation and
  {E}volution of the {G}alaxy [{T}he {G}{A}{I}{A} {S}tudy {R}eport
  ({E}{S}{A}-{S}{C}{I}(2000)4)],'' (July 2000).

\bibitem{2010SPIE.7731E..35D}
{de Bruijne}, J., {Kohley}, R., and {Prusti}, T., ``{Gaia: 1,000 million stars
  with 100 CCD detectors},'' in [{\em Society of Photo-Optical Instrumentation
  Engineers (SPIE) Conference Series}{\nolinebreak\hspace{0.1em}]},  {\em
  Society of Photo-Optical Instrumentation Engineers (SPIE) Conference Series}
  {\bf 7731} (July 2010).

\bibitem{2009SPIE.7439E..29V}
{Vosteen}, L.~L.~A., {Draaisma}, F., {van Werkhoven}, W.~P., {van Riel},
  L.~J.~M., {Mol}, M.~H., and {den Ouden}, G., ``{Wavefront sensor for the
  ESA-GAIA mission},'' in [{\em Society of Photo-Optical Instrumentation
  Engineers (SPIE) Conference Series}{\nolinebreak\hspace{0.1em}]},  {\em
  Society of Photo-Optical Instrumentation Engineers (SPIE) Conference Series}
  {\bf 7439} (Aug. 2009).

\bibitem{1978moas.coll..197L}
{Lindegren}, L., ``{Photoelectric astrometry - A comparison of methods for
  precise image location},'' in [{\em IAU Colloq. 48: Modern
  Astrometry}{\nolinebreak\hspace{0.1em}]},  {Prochazka}, F.~V. and {Tucker},
  R.~H., eds.,  197--217 (1978).

\bibitem{LL:2004BASNOCODE}
Bastian, U., ``{T}he maximum reachable astrometric precision - {T}he
  {C}ramer-{R}ao {L}imit,'' Gaia DPAC public document 2004BASNOCODE (April
  2004).

\bibitem{2010ISSIR...9..279L}
{Lindegren}, L., ``{High-accuracy positioning: astrometry},'' {\em ISSI
  Scientific Reports Series}~{\bf 9},  279--291 (2010).

\bibitem{LL:LL-078}
Lindegren, L., ``{A} general {M}aximum-{L}ikelihood algorithm for model fitting
  to {C}{C}{D} sample data,'' Gaia DPAC public document GAIA-C3-TN-LU-LL-078
  (November 2008).

\bibitem{LL:FM-030}
``{D}{P}{A}{C}: {P}roposal for the {G}aia {D}ata {P}rocessing,'' Gaia DPAC
  public document GAIA-CD-SP-DPAC-FM-030-02 (April 2007).

\end{thebibliography}
\bibliographystyle{spiebib}   

\end{document}